# A Series of Plasma Innovation Technologies by Double Glow Discharge Phenomenon

Zhong Xu[1, a)], Hongyan Wu[2, a)], Zaifeng Xu[1], Xiaoping Liu[1], Jun Huang[3]

[1] Research Institute of Surface Engineering, Taiyuan University of Technology, Taiyuan 030024, China.
[2] Department of Material Physics, Nanjing University of Information Science and Technology, Nanjing 210044, China
[3] School of Materials Science and Engineering, Nanchang Hangkong University, Nanchang 330063, China

[a)] Authors to whom correspondence should be addressed: wuhy2009@nuist.edu.cn

## Abstract

In order to break the limitation of plasma nitriding technology, which can be applied to a few non-metallic gaseous elements, the "Double Glow Discharge Phenomenon" was found and then invented the "Double Glow Plasma Surface Metallurgy Technology". This double glow plasma surface metallurgy technology can use any element in the periodic table of chemical elements for surface alloying of metal materials. Countless surface alloys with special physical and chemical properties have been produced on the surfaces of conductive materials.

By using double glow discharge phenomenon, a series of new plasma technologies，such as the double glow plasma grapheme technology, double glow plasma brazing technology, double glow plasma sintering technology, double glow plasma nanotechnology, double glow plasma cleaning technology, double glow plasma carburizing without hydrogen and so on, have been invented.

A very simple phenomenon of double glow discharge can generate about 10 plasma innovation technologies, which fully shows that there is still a lot of innovation space on the basis of classical physics.

This paper briefly introduces the basic principles, functions and characteristics of each technology. The application prospects and development directions of plasma in metal materials and machinery manufacturing industry will also be discussed.

## 1. Introduction

In 1978, on the basis of many years' research on plasma nitriding technology [1-8], we were trying to break the limitation that ion nitriding technology can only be applied to a few gaseous non-metallic elements. In order to obtain the surface alloying under glow discharge condition by using solid metal elements, we came to understand that the key problem is how to vaporize the solid metal elements, so that the glow discharge gas includes the atoms of the solid metal elements.

In the plasma nitriding experiment, we observed that a spark and/or local micro arc discharge emitted from the surface of the working piece. In addition, we also observed that there are more and more steel fine powders on the stove chassis. It had made us to realize that, under the glow discharge condition, the solid metal elements in cathode electrode may be sputtered by positive ion bombardment from plasma and become gas metal element.

This observation made us realize that we can use the ion bombardment and sputtering to achieve gasification of solid alloy elements. We set up a second cathode (as a source electrode, made of the desired alloying elements) between the anode and cathode in plasma nitriding equipment. Driven by two DC power supplies, two glow discharge (plasma) zones can be established respectively between the anode and cathode as well as the anode and source cathode. We called this state as "Double Glow Discharge Phenomenon". Ion bombardment at the source cathode makes the desired solid alloying element to be sputtered and gasified into glow discharge space.

In 1980, we invented the "Double Glow Plasma Surface Alloying Technology" based on the "Double Glow Discharge Phenomenon"[9]. We discovered that the Double Glow Plasma Surface Alloying Technology can be applied to any solid chemical element such as tungsten, molybdenum, nickel, chromium, etc. and their combination to conduct surface alloying. Surface alloys with gradient concentration of alloying elements have been produced on the surfaces of steels, titanium alloys and inter-metallic compounds, etc.

Inspired by the double glow plasma surface metallurgy technology, a series of other new plasma technologies, such as the double glow plasma grapheme technology, the double glow plasma brazing technology, the double glow plasma sintering technology, the double glow plasma nanotechnology, the double glow plasma cleaning technology, the double glow plama carburizing without hydrogen and so on, have been invented.

## 2. Basic Background Concepts

### 2.1 Plasma

The plasma is a neutral ionized gas consisting of electrons, positive ions and neutral atoms. Plasma may be generated by the passage of a current between electrodes, by induction, or by a combination of other methods. In 1928, American scientists Irving Langmuir first introduced the term "plasma" into physics [10].

As we all know, with temperature increase, there will be a solid, liquid and gas three-state transition process in the matter of nature. As the temperature rises further, a large number of outer electrons of an atom will break away from the bondage of the nucleus and become free electrons, and the atoms that lose electrons will become positive ions. This process is called ionization, and the result is the transformation of matter into plasma.

The plasma is called the fourth state of matter. Plasmas are very wide spread in the universe. According to the calculations of M. Saha, an Indian astro-physicist, almost 99.9% of matter exists in plasma state throughout the universe. For example, from burning fires to the sun that breeds all things, from beautiful aurora to glittering galaxies. Stars and interplanetary space consist of plasma. Plasmas can also be produced by artificial methods, such as nuclear fusion, fission, glow discharge and arc discharge.

In addition, plasma is a kind of gas material with high energy. The plasma is very active, has very good conductivity, and has strong interaction with other substances and electromagnetic fields. Therefore, the plasma will belay an increasingly important role in modern industry, agriculture and other various fields.

The plasma is the most important form of matter in the universe. With the deepening and development of human scientific research in nature, the research of new technologies in the fields of universe, celestial bodies, space-time, satellite, space flight, energy and so on will enter a new era of plasma research.

**2.2 Glow discharge**
In 1775, American scientist B. Franklin was the first person to study lightning in the atmosphere. After many experiments, it has been proved that lightning in the sky is a discharge phenomenon. In 1835, M. Faraday discovered glow discharge and Faraday dark area when he studied low voltage discharge [11].

Glow discharge can be is produced by placing two parallel electrodes in a closed low vacuum container and applying a DC power supply between the two electrodes. Electrons produced in discharge collide with gaseous atoms or molecules, exciting and ionizing gaseous atoms. When stimulated electrons fall back from the excited state to the ground state, they release energy in the form of light.

Glow discharge refers to the phenomenon of glow discharge in low pressure gas, that is, self-sustaining discharge in rarefied gas. Its discharge parameters are characterized by high voltage and low current density.

The main application of glow discharge is to use its luminous effect, such as neon lamp and fluorescent lamp [12-15]. Glow discharge is also used in composition analysis of material surface peeling, surface purification, sewage treatment, sterilization and disinfection, ion source of analytical instruments, plasma nitriding, double glow surface alloying, etc [16-21].

**2.3 Arc Discharge**
Arc discharge refers to the phenomenon of gas discharge with arc white light and high temperature. Usually, the method of arc discharge is to separate the two electrodes immediately after contacting. Because of short circuit heating, the cathode surface temperature increases sharply and hot electron emission occurs. Thermo-electron emission makes collision ionization and secondary electron emission of cathode increase sharply, which makes the gas between the two poles have good conductivity. In the case of arc discharge, the voltage between the two poles decreases with the increase of current, which has a strong arc. Its discharge parameters are characterized by low voltage and high current density.

Arc discharge is the most widely used industrial application is arc welding [22-25]. Especially in the machinery manufacturing industry, ship manufacturing, automobile manufacturing, high-rise buildings, large structural parts and so on, are indispensable processes. In addition, Arc discharge can be used as a strong light source, as a light source for exciting element spectrum in spectral analysis, as a light source for smelting, welding and cutting of high melting point metals in industry, as an ultraviolet source (mercury arc lamp) in medicine, and so on. In recent years, arc discharge has been used as a source of metal elements for surface modification technologies such as physical vapor deposition.

**2.4 Difference of Glow Discharge and Arc Discharge**

The most important difference between glow discharge and arc discharge is that glow discharge is an area discharge, while arc discharge is a point discharge. Glow discharge is dispersed and stable, and arc discharge is concentrated and intense. The high current of arc discharge concentrates at one point, which will lead to a sharp increase in temperature of the work-piece and cause burning and damage of the work-piece. Therefore, arc discharge must be prevented in the process of plasma nitriding and plasma surface alloying; otherwise, arc discharge may lead to melting of work-piece and damage of equipment.

**2.5 Double glow discharge and its hollow cathode discharge phenomenon**

In order to break the limitation that plasma nitriding can only be applied to non-metallic elements, we found the phenomenon of double glow discharge as shown as Figure 1(a) in 1978. The Double Glow discharge is formed by two sets of cathodes (i.e., the work-piece and the source) driven with two different electric potentials. Experimental apparatus for the formation of double glow discharge and its hollow cathode discharge in is shown in Figure 1.

The experimental device is set in a sealed vacuum container, including the first cathode (Z2) and the 2$^{nd}$ cathode (Z1) (as is known as the source electrode), two power supplies (V1 and V2). Two cathodes made of a low-carbon steel plate are placed in parallel with a relative distance adjustable in the range of 10-l00mm. The supply output voltage can range of 0-1000V. Working discharge gas is industrial-grade pure argon and working gas pressure ranges in 10-130Pa.

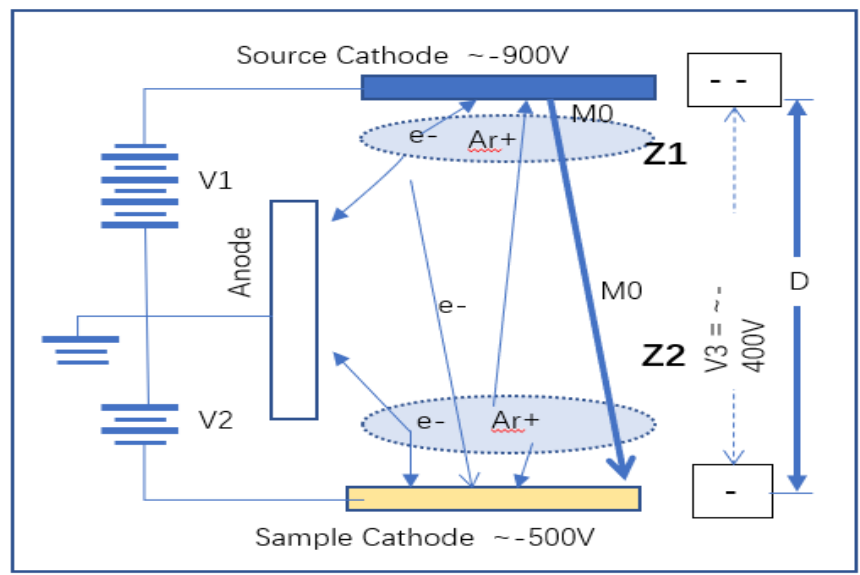

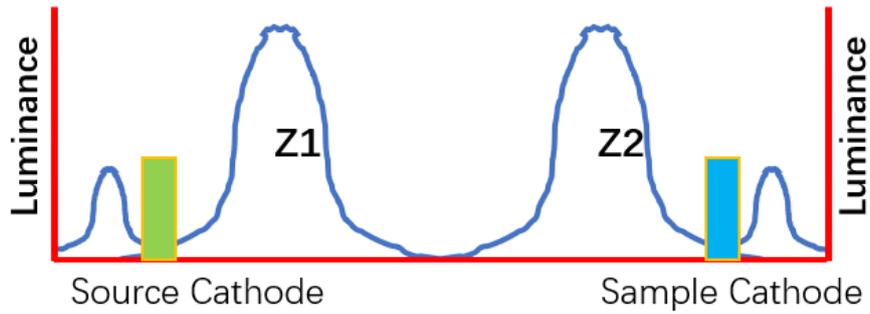

(a)

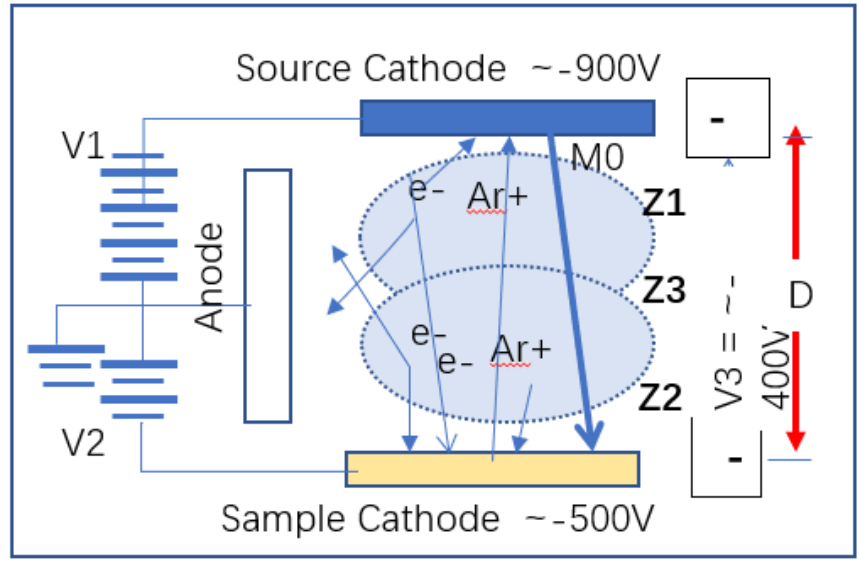

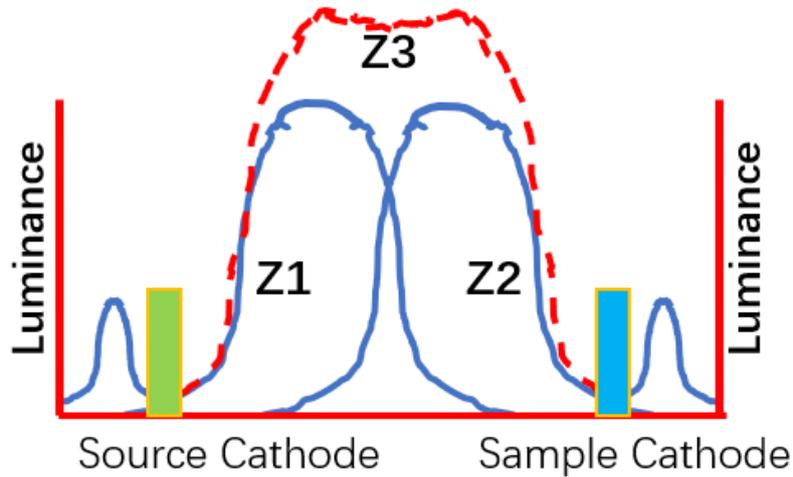

(b)

FIG. 1. Experiment setup for double glow discharge, (a) intensity of double glow discharge. Z1 and Z2 on Source and Samples cathodes and no intersection between source and sample. (b) Z3 is the total glow discharge intensity when two cathodes get closed and the glow discharge zones overlap.

When two power supplies (V1 and V2) are switched on, the glow discharges are generated along the surface of cathode and sample respectively. At first, the negative glow regions of the two cathodes ~~is~~ are well defined and do not intersect as shown by the curve Z1and Z2 in Figure 1(a). Double glow discharge can also produce hollow cathode discharge as shown as Figure 1(b). As argon pressure decreases, the thickness of the negative glow region increases. When the negative glow regions of the source Z1 and the sample Z2 are overlapped with each other in the space between two cathodes, the glow brightness is significantly enhanced. If the pressure is reduced further or the two cathode voltages are increased, the two cathode glow regions are mutual overlap and cross, the brightness of the glow discharge and two cathodes current density will greatly increase shown as curve Z3. This is the Double Glow Hollow Cathode Discharge (DG-HCD). Since the discharge potentials of the two cathodes are not equal, we call this phenomenon also as the unequal potential hollow cathode discharge.

The current amplification effect of the double glow hollow cathode discharge is shown in Figure 2 [5]. When the cathode voltage $U_c$ is increasing to 400V, then both the sample cathode current $I_c$ and source electrode current $I_s$ will sharply increase.

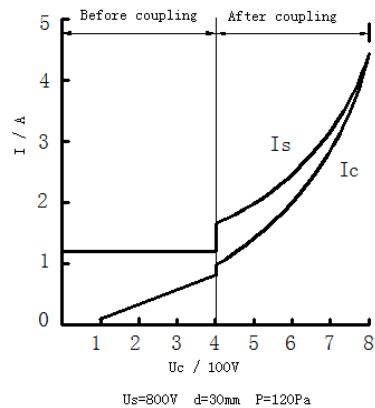

FIG. 2. Current amplification effect of unequal DG-HCD

The luminous intensity and current density of double glow hollow cathode discharge can be 1 to 3 orders of magnitude higher than that of common glow discharge.

## 3. Double Glow Plasma Graphene Technology

Graphene, as a one-atom-thick layer of carbon, has been extensively studied since its first discovery in 2004, and they have shown great application prospects in many fields, such as photodetector, conductive switching, bio-imaging, etc. A great challenge in the applications of graphene is the preparation of a large-scale graphene film grown on arbitrary substrates. Although there have been much progresses in fabricating graphene, such as mechanical exfoliation [26], epitaxial growth [27], chemical vapor deposition [28] and reduction of graphene oxide (GO) [29], the cost, quality and controllability of the graphene still remain with issues. In addition, it is difficult to accomplish one-stage grow large-area graphene films on the demanded substrate, avoiding from the transfers of graphene at the expense of the loss of Ni or Cu. Simultaneously, the good adhesion between the graphene film and the various substrate is required. The Xu-Tec has the capability of enhancing the sputtering yield through the principle of the double-glow plasma discharge.

Figure 3(a) showed the schematic diagram of the Xu-Tec device. The double-glow discharges occur in the vacuum chamber. One glow discharge produced by the workpiece electrode heated the substrate (the quartz substrate). The other glow from the target-electrode discharge (graphene-oxide target material) produced the needed graphene film. The cross-linked reactions between double-glow Ar+ plasma enhanced the bombardment energy for the target elements and substrate surface, so that the graphene-oxide target can be bombarded a piece of graphene patches to form GO films, shown in Figure 3(b). Due to the strong bombardment for the GO sheets, crumpled GO films were prepared on the quartz matrix. Based on the mechanisms of the double-glow plasma bombardment, GO films can form on the arbitrary substrate. In addition, because the surface of the substrate was bombarded to produce amounts of defects, there is the excellent adhesion between the graphene nanosheets and the quartz substrate. In the previous research [30], the growth processing of graphene-oxide film was obtained. Graphene paper (Dimensions as required, purity 99.9%) was used as sputtering targets under the source voltage of 900 V. The tank is evacuated to a pressure of about $5\times10^{-4}$ Pa and then filled with argon (99.999%) to 30 Pa. The quartz plates (30 mm×30 mm×2 mm) as the substrate were sequentially cleaned ultrasonically by acetone at room temperature and then the substrate was connected with the work-piece electrode at the working voltage of 400V. The working time was set for 20 min at the temperature of 600 ℃, and then samples were cooled to room temperature in the vacuum chamber.

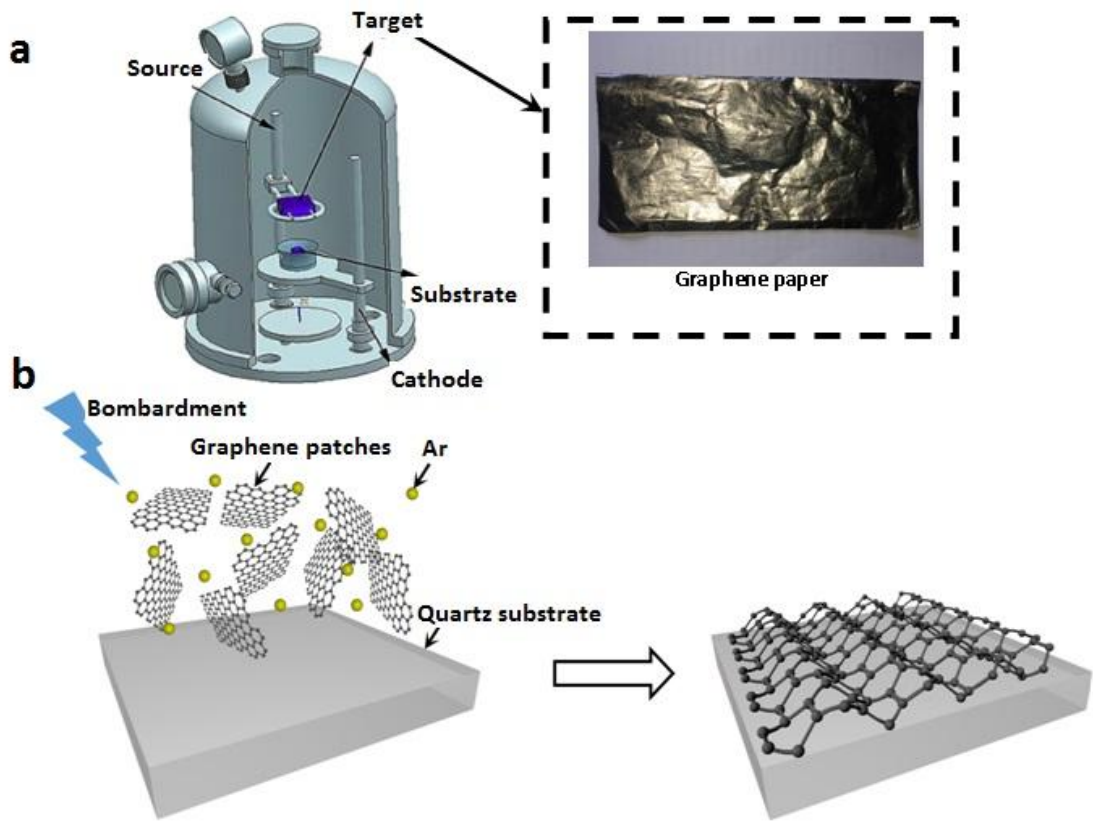

FIG. 3 (a) The schematic diagram of the Xu-Tec device and (b) the growth procedures of GO film under the double-glow plasma bombardment [44].

As-prepared graphene film prepared using Xu-Tec process exhibits the special structural feature, which can be developed in the future. Firstly, the as-prepared graphene displays the morphology in a form of nanosheets structure, shown in Figure 4(a) and the amounts of wrinkles and folding were present on the graphene nanosheets. The presence of the crumpled graphene was beneficial for the photoresponsivity of a graphene-based photodetector, which is consistent with the literature report [31]. The as-prepared graphene with different defect states exhibits the special UV-induced photoresponsivity. Raman spectra of GO film reveals two prominent bands at 1346 cm$^{-1}$ (D bands) and 1608 cm$^{-1}$ (G bands) before and after UV irradiation (Figure 5 (b)). After UV treatment, the $I_D/I_G$ ratio decreased from 0.89 to 0.82. This can be assigned to formation of further sp$^2$ bonds after the UV-assisted reduction of the graphene oxide sheet [32]. Additionally, 2D band peak at 2608 cm$^{-1}$ suddenly appeared, which indicates a reduction of the GO film. The surface structure and composition of GO film were greatly influenced by the UV irradiation. Therefore, the as-prepared graphene film due to the presence of wrinkles and folding will exhibit high UV-induced photoresponsivity.

In addition, we also attempted to change the target materials. The original graphene-paper target was replaced by the graphite paper. The surprising result is that the similar GO films form on the substrate. The higher density of D peak display that the GO film prepared by graphite paper exhibits more edge defects than that of prepared by graphene-paper target, shown in Figure 5. The decrease in density of 2D peak indicates that few-layer GO films can be prepared by the above target materials. Further research is proceeding. We aim to prepare the defect-controlled GO films by the adjusting the plasma processing.

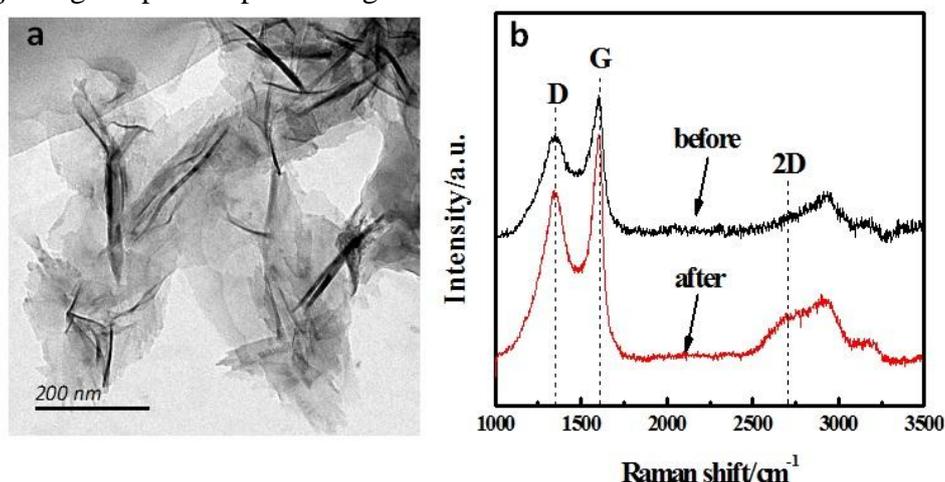

FIG. 4. (a) TEM image typically of graphene nanosheets, which was exfoliated from the as-prepared GO film on the quartz substrate. (b) Raman spectra of the GO film before and after UV irradiation.

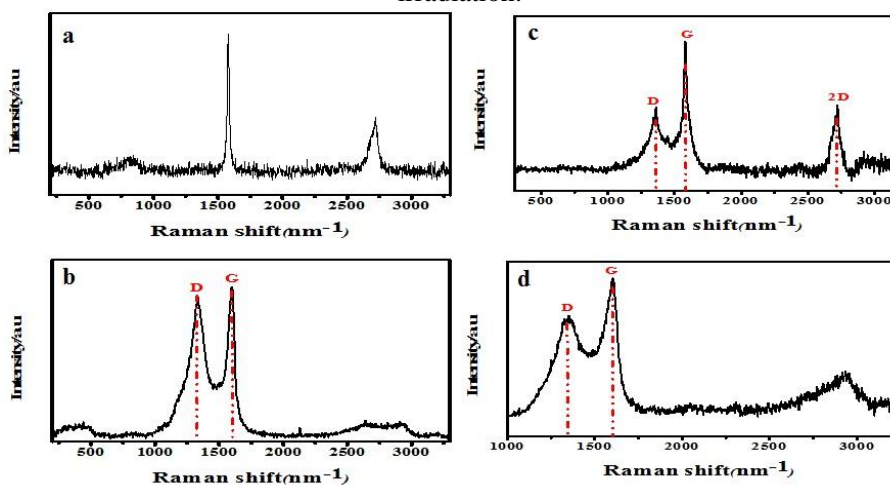

FIG. 5. Raman spectra of (a) and (b) graphit-paper target and the graphene oxide film on the substrate, (c) and (d) graphene-paper target and the graphene oxide film on the substrate

We've concluded that the Xu-Tec process can be used as a new potential technology for large-area preparation of graphene and its composite films. The advantages are as follows:

(1) The target material (graphene-oxide paper, graphite or other target materials) can be designed the required sizes of the substrate and there is no limit to the component of the substrate.

(2) The cross-linked reactions of double-glow $Ar^+$ plasma enhance the bombardment energy between the target elements and substrate surface, so that the target materials can be bombarded under the high sputtering energy.

(3) The as-prepared graphene film can be directly formed on the different substrate (e.g. quartz or metal surface), which avoids the transfers of graphene at the expense of the loss of Ni or Cu and simple physical coating of GO powder.

(4) The excellent adhesion between the graphene film and substrate is confirmed. The graphene film and its composite films can be prepared on the demanded substrate. The introduction of crumpled graphene to composite film could provide an effective transmission channel for photogenerated electrons and weaken the recombination of charge carriers.

Although the as-prepared graphene exhibits the presence of some defects to a certain extent, it is essential for graphene to fulfill functional properties. We believe that the Double Glow Plasma Technique provide a new economic and simple approach to prepare the large-area graphene and its composite films.

## 4. Double Glow Plasma Brazing Technology

The double glow plasma surface metallurgy device is not only used for surface alloying of materials, but also for other technologies. The double glow brazing technology [33] is a typical example.

Vacuum brazing is a common method to connect thin-walled sections and precise parts made of special materials. Vacuum brazing requires high vacuum, such as brazing of aluminium waveguide tubes in large radar equipment. The vacuum brazing furnace required not only super large vacuum chamber size, but also strict vacuum and temperature control accuracy. Its special equipment is very expensive. The operation of such a large vacuum equipment has high energy consumption and low thermal efficiency. Aluminium and its alloys tend to form very stable oxide film on the surface, which deteriorates the wettability of brazing filler metal. If it can not be effectively removed in brazing operation, it will lead to low welding strength and even virtual welding. Therefore, special surface cleaning must be done before brazing of aluminium and its alloys. In addition to brazing operation under high vacuum, activating agent must be added to the flux or magnesium vapor must be introduced into the furnace during brazing. In this way, it is easy to cause damage to vacuum equipment, and the workpiece must be thoroughly cleaned after welding.

Using double glow plasma surface metallurgy equipment and using rare gas glow discharge as heating source for brazing operation can effectively solve this shortcoming of the traditional process. Figure 6 is a schematic diagram of the working principle of double glow brazing technology. The workpiece is connected with the cathode. The auxiliary cathode and the additional cathode are arranged according to the requirement of hollow cathode effect. In this way, hollow cathode discharge effect is produced on the surface of glow discharge and part of the workpiece. Intense ion bombardment has a high energy density, so heating the workpiece has a high thermal efficiency. In addition, the working pressure during brazing operation ranges from 10 to 150 Pa, which reduces the requirement of vacuum system. Compared with traditional vacuum brazing furnace, the energy saving effect is significant. At the same time, the continuous bombardment of positive ions on the workpiece surface has a high efficiency of surface cleaning and purification. Therefore, using plasma as heating source for brazing can obtain better welding quality, or can reduce the amount of flux added in solder, or even achieve no cleaning after welding.

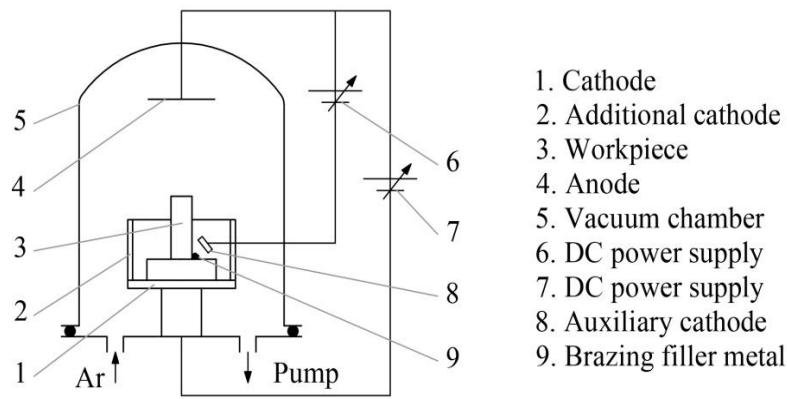

FIG. 6. Schematics of brazing process setup with the double glow plasma technology

Double glow brazing technology has been successfully applied in the multifunction parts. The diameter of the parts is 60mm, the wall thickness is 0.3mm, the material is 4J36, and the brazing metal is AgCu28. The workpiece after routine cleaning is loaded into the furnace and filled with brazing filler metal, vacuum is pumped to 4Pa, then filled with argon and heated by glow discharge. The workpiece will soon reach the brazing process temperature. By adjusting the parameters of DC power supply, the brazing filler metal will be fully filled and the gas pressure and current will then be reduced. The brazing filler metal will be solidified before it is discharged from the furnace.

The greatest advantage of double glow brazing technology is that it does not need to heat the brazed components as a whole, but to heat the brazed parts locally. This can not only greatly shorten the brazing time, but also greatly reduce the workpiece deformation. As usual, vacuum brazing lasts at least 6 to 8 hours, while double glow brazing lasts less than 3 hours. This not only reduces the energy consumption, but also greatly improves the production efficiency.

The advantages of double glow plasma brazing technology are as follows:
(1) Lower cost of vacuum requirement, the working pressure in a range of 10~150 Pa for brazing operation.
(2) Under glow discharge conditions, ion bombardment will remove oxides and other contaminants on the surface of work-pieces and brazing materials, which can greatly improve the quality and reliability of welding.
(3) By using double glow discharge, only the localized heating of the work-piece is needed. This can greatly reduce the deformation of the work-piece.
(4) Local heating can greatly shorten the processing time, improve efficiency and reduce costs.

# 5. Double Glow Plasma Sintering Technology

For the production of powder metallurgical products, sintering is a key process. The final mechanical properties and geometric accuracy of products are closely related to the selection and control of sintering process parameters. The recent technological development trend of this industry is the continuous expansion of sintering materials. Many oxidizing metals such as aluminium, titanium and other parts have been processed by powder metallurgy process in order to reduce manufacturing costs. In such cases, the sintering atmosphere is demanded. The mechanical and physical properties (especially strength) of products are also required to be close to the dense materials. Therefore, high temperature sintering and supersolid line sintering become more and more of choices. For example, for high strength iron-based structural parts, the sintering temperature of 1300-1350$^o$C is much higher than 1100$^o$C of traditional process, which brings higher requirements for sintering equipment.

At present, the sintering heating method widely used in production is still resistance heating [34]. In addition, flame heating, induction heating and direct low-voltage high-current heating (vertical melting method) are also used in some specific situations. [35-37] Resistance heating elements include resistance wires, silicon carbon rods, tungsten and molybdenum wires, etc. Domestic ordinary sintering furnace mostly uses ferrochrome-aluminum resistance wire, whose long-term working temperature cannot exceed 1150$^o$C. If silicon-carbon rod is used, the working temperature can be raised to 1350$^o$C, but it is easy to age and has limited service life. Tungsten wire or molybdenum wire can significantly increase the heating temperature, but it is limited by the high temperature performance of refractory and insulating materials, and the working temperature generally does not exceed 2000 $^o$C. Flame heating can also not achieve high temperature, while induction heating can achieve a high temperature of 3000$^o$C, but induction generation equipment is expensive, power factor is low and it is not easy to obtain high productivity. Vertical melting is a sintering technology specially used to produce refractory metals such as tungsten and molybdenum products, but this method is only suitable for rod products with equal cross-section area, and has considerable energy consumption. Therefore, the development of flexible and reliable high temperature sintering methods has become a research focus in order to save energy or meet the needs of technological development. In this context, a variety of new dense energy sintering technologies has emerged, of which the typical ones are microwave sintering and plasma sintering. Because plasma has the unique effect of reducing and activating the surface of powders, it has attracted more and more attention in the sintering applications of oxidizing metals, such as titanium and its alloy products.

Double glow plasma sintering technology is a novel technology of energy intensified high-temperature sintering process. The schematic diagram of this technology is shown in Figure 7.

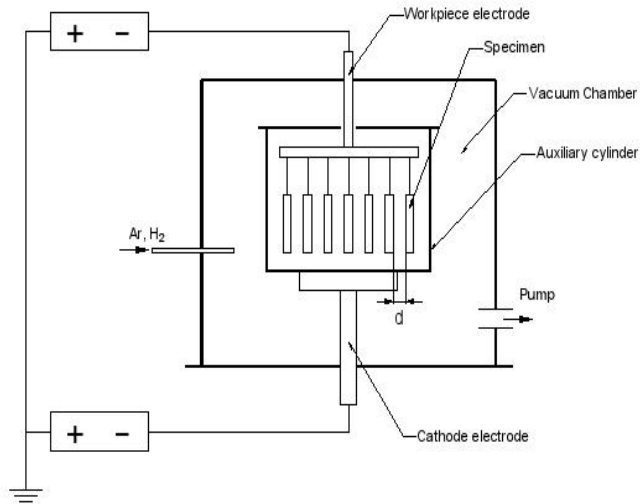

FIG. 7. Schematic Diagram of double glow plasma sintering

Double glow sintering technology [38] derived from double glow plasma surface metallurgy technology is a new high temperature sintering technology with high efficiency and reliability. In the traditional double glow discharge device, the blank to be sintered is placed on the cathode, and strong hollow cathode discharge effect is produced by designing the cathode structure and adding auxiliary source. In the case of hollow cathode discharge, the positive ions are accelerated in the cathode drop area and bombarded the billet surface with high energy. The energy of these ions, apart from a small part used to produce cathode sputtering, is mostly converted into heat, which causes the latter to sinter at a rapid rise in temperature.

From the aspects of heating principle, physical process of ion bombardment in sintering and process control, double glow sintering technology has the following characteristics:
(1) The thermal efficiency of the whole sintering process is very high because of the heating method of hollow cathode discharge.
(2) Higher sintering temperature (above 3000$^o$C) can be obtained.
(3) Ion bombardment on the surface of sintered billet has significant activation sintering effect.
(4) Chemical heat treatment such as carburizing and nitriding can be realized at the same time of sintering by introducing working atmosphere, or auxiliary cathode can be introduced to infiltrate metals such as Cr, Mo and Ni on the surface of sintered billet at the same time.
Continuous temperature regulation and accurate control can be realized by changing the parameters of DC power supply.

Double glow sintering technology can be widely used to sinter metals, alloys, ceramics, and graphite products. Especially it can be treated by chemical heat treatment or surface alloying at the same time. It has advantages in manufacturing high wear-resistant and heat-resistant parts, and can even be used to manufacture functionally gradient materials, showing great application value and prospects. Fig. 10 is an application example of sintered tungsten rod. Reflective screen is installed in the furnace. The distance d between workpieces should be set in the range of 8-20 mm, and the sintering temperature can reach 2800-3000$^o$C.

# 6. Double Glow Plasma Nano-Powder Technology

The preparation of nano-powders can be divided into gas phase method and liquid phase method. The gas phase method includes chemical vapor deposition (CVD), laser chemical vapor deposition (LCVD), vacuum vapor and electron beam sputtering, etc [39-41]. The disadvantage is that the equipment requirements are high, hence the investment is large. The liquid phase method includes sol-gel (SOL-GEL) method, hydrothermal method (hydrothermal synthesis) and precipitation (co-precipitation). Among them, Sol-Gel has been widely used, the main reasons are as follows:

1) Simple operation: short processing time, no need for extreme conditions and complex instruments and equipment;

2) Molecular mixing of each component in solution, which can prepare various nano-powders with complex and large distribution;

3) Strong adaptability: it is convenient to prepare micro-powders. It is also used to prepare fibers, thin films, porous carriers and composite materials.

In the process of double glow ion discharge, large numbers of atoms, ions and atomic clusters are sputtered from the source electrode. These tiny particles are nano-sized. If appropriate condensation and collection are carried out, nano-powders with source material composition can be obtained. Based on this fact, the technology of preparing nanomaterials by double glow hollow cathode discharge has been derived by modifying common double glow discharge devices and adding chillers and collecting devices [42].

Figure 8 is a schematic diagram of the working principle of preparing nanomaterials by double glow hollow cathode discharge. In order to obtain the maximum sputtering amount, the bulk material to be prepared into nano-powders is placed in the supply cathode. In order to obtain the maximum sputtering amount, the supply cathode must be made into a hollow structure to produce strong hollow cathode discharge effect. The distance between them is 5-30 mm. The collection container is placed under the supply cathode as another cathode, and it itself has strong cooling. Generally cooling uses water or liquid nitrogen. Chiller with and furnace wall are grounded as anode. The supply voltage of cathode is controlled in the range from -800V to -1500V, while the collector voltage is adjusted from -100 to -500V. Under these conditions, the supply cathode produces intense hollow cathode discharge, and a large number of atoms, atomic clusters and ions are sputtered by the field emission mechanism. These particles migrate rapidly to the collector cathode under bias pressure, and then cooled rapidly by chiller, and finally deposited in the collector as solid nanoparticles.

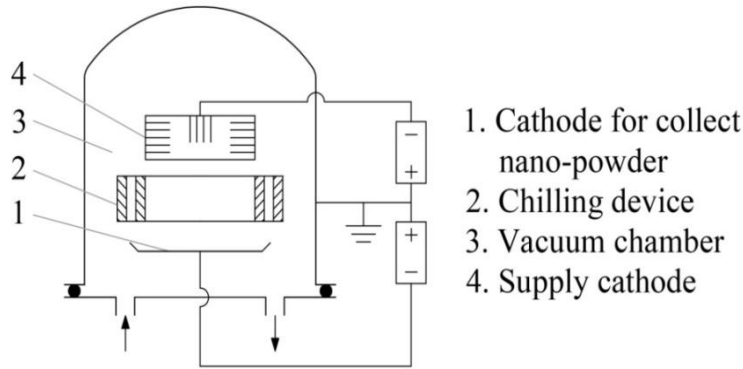

FIG. 8. Schematic of nano-materials production by double glow plasma sputtering

The technology of preparing nanomaterials by double glow hollow cathode discharge has the following advantages:

(1) The supply source can be pure metals or alloys prepared by metallurgical method or mixtures of metal powders prepared by powder metallurgy method in any proportion, so that pure metals, alloys or multi-element nanomaterials can be obtained.

(2) Non-metallic materials can be used as cathodes and made into nanomaterials as long as they are conductive.

(3) Atomic-scale particles sputtered from the supply cathode are smaller than those obtained by conventional methods.

(4) Since supply cathode can be made of only one kind of material, and the preparation process has no pollution source, so its nanometer material has high purity.

(5) No emissions or pollution.

## 7. Double Glow Plasma Sputter Cleaning Technology

To prevent the inner surface of tubes/pipes from corrosion, oxidation, erosion or wear, a functional coating is typically applied. The coating may be a layer of metallic or ceramic coating. For instance, to increase the erosion/corrosion resistance of gun barrels for tanks, Cr and Ta coatings (100~200μm) may be deposited.

In order to increase the coating adhesion to the substrate, and hence extend the coating lifetime, it is useful to clean its surface before coating deposition. One may consider chemical or plasma cleaning. Chemical cleaning is to remove the dirt, rust or oil/grease from the surface, and it may be performed before the substrate enters the vacuum deposition system. Plasma cleaning is typically designed to remove the surface oxide ("native oxide") which exists on nearly all metal surfaces when they are exposed to air and other contaminates that cannot be easily removed using chemical cleaning.

Figure 9 illustrates the device for providing double glow discharge [43]. When two parallel plates are placed in a vacuum chamber, which has been evacuated (e.g. to $1\times10^{-5}$ ton) followed by introduction of an inert gas argon to a pressure of 150 mTorr, and if the two plates are biased with negative voltage as illustrated, a double glow discharge can be formed. The double glow discharge may be more specifically used for cleaning a substrate surface A when the voltage on plate A may be a negative DC voltage of -600V to -1500V. The voltage on the other plate B may also be a negative DC voltage of -100V to -200V. In this manner, sputter cleaning of plate A will take place, thereby remove oxides and other surface components that may interfere with an ensuing coating process.

One should note that such plasma is relatively intense and a high current can be obtained from both power supplies.

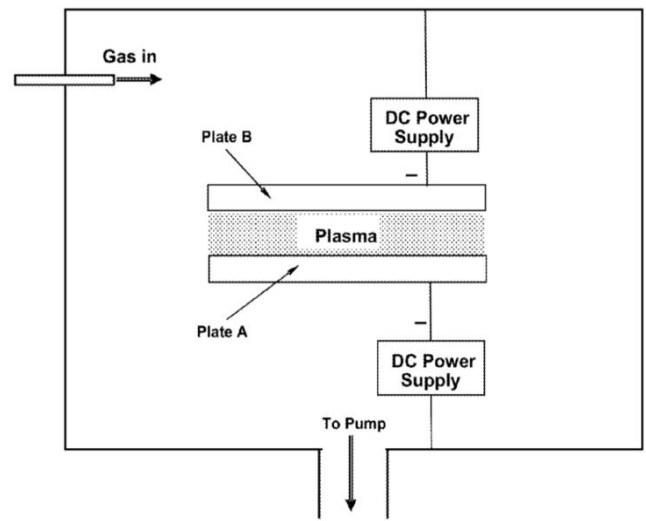

FIG. 9. Schematic for plasma cleaning by double glow discharge

Due to high energy and high density plasma produced by double glow discharge, double glow plasma sputter cleaning has the advantages of quick and thorough cleaning and purification. Dr. Wei et al in Southwest Research Institute, as disclosed in a USA patent [43] demonstrate one application example for gun tube. A shield tube is provided as the second cathode inside a gun tube. Two negative bias power supplies are applied with a higher voltage for the gun tube and a low one for the shield tube. Hollow glow discharge plasma is generated in the space between two tubes. Sputter cleaning takes place on the inner surface on the gun tube.

## 8. Double Glow Plasma Carburizing without Hydrogen

Double glow plasma surface metallurgy can use solid graphite as a carbon source. Active carbon atoms produced by ion sputtering deposit and diffuse on the surface of titanium sample, thus plasma carburizing under non-hydrogen environment is realized. In the same way, the non-hydrogen carbonitriding can also be realized if a mixture atmosphere of argon and nitrogen is used as discharging gas. [44-47].

Figure 10 shows the sectional morphology of non-hydrogen carburizing layer of industrial pure titanium under different carburizing temperatures, time and gas partial pressures, respectively. As the carburizing temperature exceeds the transformation temperature of titanium, the phase structure has transformed from α to β. When the temperature is higher than $1000^oC$, the structure turns into Widmannstatten structure. The pure titanium after non-hydrogen carburizing exhibits the high value of 1400HK at the surface due to the formation of high hardness TiC and the surface hardening consists of dislocation hardening, solid solution hardening and so on as shown in Figure 10. In addition to high hardness, the wear rates of carburized samples are about 2%-7% of that of the original one. This shows the wear rate can be reduced by carburizing. Due to the TiC wear resistant phase formed and solid solution strengthen effect, the wear resistance of TC4 can be-significantly improved, and the wear rate is much reduced.

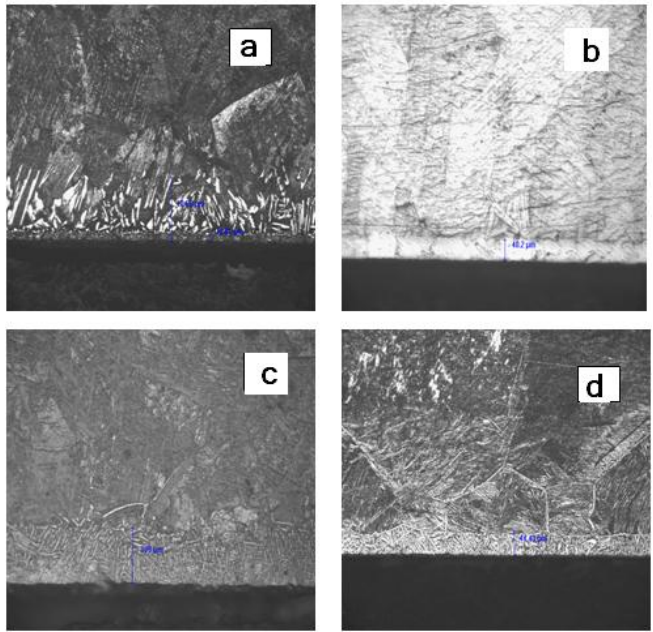

FIG. 10. Sectional microsturcture of pure titanium after carburizing (a) 1000°C, 3h, 100×; (b) 980°C, 2h, 200×; (c) 960°C, 3h, 200×; (d) 980°C, 2.4h, 200×.

Figure 11 shows a hardness curve of pure titanium after non-hydrogen carburizing. The hardness as a function of depth is a parabolic-shaped curve, with the maximum value at the surface. Besides the formation of high hardness TiC, the mechanism of surface hardening consists of dislocation hardening, solid solution hardening and so on.

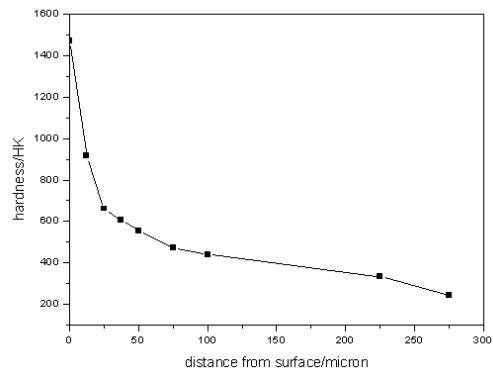

FIG. 11. Hardness curve of pure titanium after non-hydrogen carburizing

The friction coefficient of a non-hydrogen carburizing layer on titanium surface was studied by a ball-disk wear test. The wear resistance of the material was characterized by specific wear rate. To determine the weight loss rate, it is required to weight the sample before and after wear test. The wear rate of original TC4 samples is 0.0934g.h$^{-1}$. The wear rates of four samples after carburization are 0.0032g h$^{-1}$, 0.0064g h$^{-1}$, 0.0026g h$^{-1}$ and 0.0019g h$^{-1}$ as shown in Figure 12. The wear rates of carburized samples are about 2%-7% of that of the original sample. It demonstrates that wear rate can be reduced by carburizing. Due to the TiC wear resistant phase formed and solid solution strengthen effect, the wear resistance of TC4 can be much improved, hence the wear rate is significantly reduced.

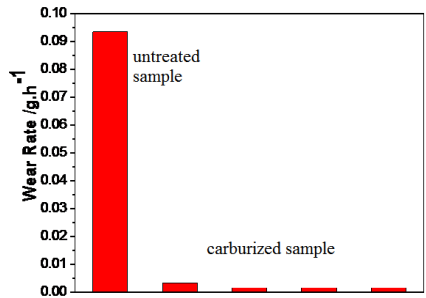
FIG. 12. Wear rate of different samples

The double glow plasma carburizing without hydrogen, has been successfully applied in industry products, such as titanium alloy valves, fasteners, hydraulic cylinders, gears, bearings [48] as well as drills, screws and nuts. These components play a key role in improving the performance of devices. Figure 13 shows some developed titanium alloy products. The application of titanium alloy formed by the Xu-Tec process has shown a very promising future.

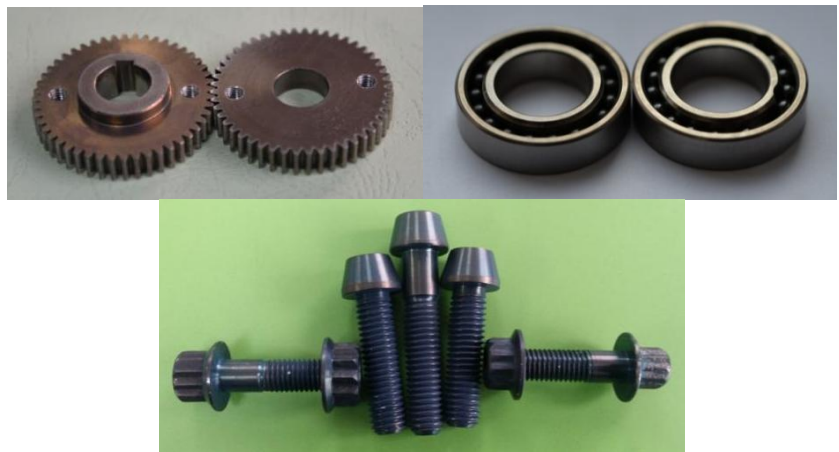
FIG. 13. Titanium gear, bearing and screw treated by Xu-Tec process

## 9. Double Glow Plus High Frequency Plasma Surface Alloying

Double glow plus high frequency plasma surface alloying technology is an energy-intensified surface metallurgy technology with high efficiency and reliability. The schematic diagram of this technology is shown in Figure 14 [49]. The double glow plus high frequency plasma surface alloying process is to set a high-frequency induction device between the cathode and the source of the double-glow plasma surface metallurgy equipment. This device is not a sputtering supply source, but a stepless discharge that generates a glow discharge ionization body, which has the function of auxiliary heating source sputtering material and cathode workpiece.

The high-frequency induction device generates strong magnetic field and electric field, which enhance the glow discharge between the cathode and the source. This increases ionization rate and provides an accelerated formation of the diffusion layer, a deposition layer and interdiffusion layer formed on the surface of the conductive workpiece.

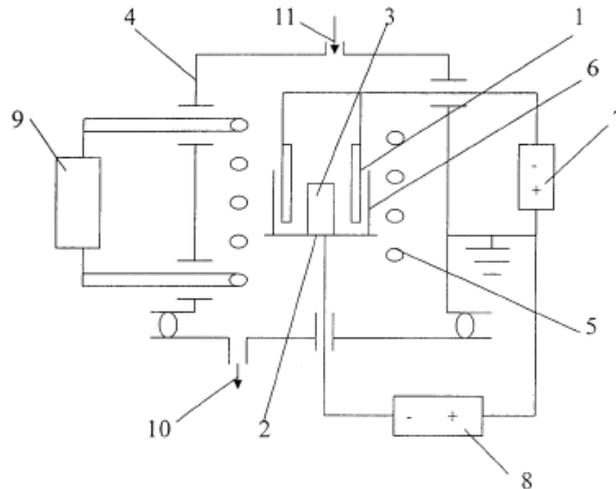

1. Source electrode, 2. Workpiece cathode, 3. Workpiece 4. Furnace cover (grounded anode), 5. High-frequency induction device, 6. Auxiliary cathode, 7. High-voltage DC sputtering power source, 8. High-voltage DC heating power source, 9. High-frequency generator, 10. Vacuum system, 11. Air intake system.
FIG. 14. Schematic of double glow plus high frequency plasma surface alloying

The technology of the double glow plus high frequency plasma surface alloying has the following advantages. The addition of the high-frequency induction device has the additional advantageous as the following.
(1) Increases the energy of the electrons, the collision probability of the electrons and `the ionization rate. The glow discharge current increases due to the effect of the strong electric and magnetic fields. The hollow cathode effect will be further strengthened. The ion current sputters out the elements in the source electrode in the form of atoms, ions or various particles.
(2) Quickly heat the source and the workpiece.
(3) Reduce the discharge working pressure by 1 to 2 orders of magnitude compared with the DC discharge working pressure, and increase the bonding force of the penetration coating.
(4) Increase the speed of infiltration and deposition.

## 10. Tri-Glow Plasma Surface Alloying

Tri-glow plasma surface metallurgy technology is the modified sputtering-supply method based on the double-glow plasma surface alloying [50]. It changes the original dual-cathode structure to three cathodes, using the hollow cathode and plume effect to generate sputtering plasma. This forms a plating layer, a diffusion layer, and a diffusion-plating composite layer on the surface of metal or non-metal solid substrates. The greatest feature of this technology is to achieve the plating of alloy elements on the surface of non-conductive materials.

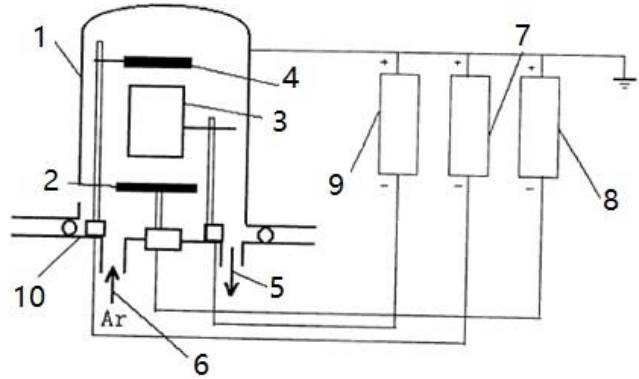

1. Furnace cover 2. S pole 3. P pole 4. W pole 5. Vacuum system 6. Working gas 7. W pole DC power source 8. S pole DC power source 9. P pole DC power supply 10. The bottom plate of the furnace

FIG. 15. Schematic of tri-glow plasma surface alloying

Figure 15 is a schematic diagram of the working principle of tri-glow plasma surface alloying in order to obtain the maximum sputtering amount and the formation of alloy elements on the surface of non-conductive materials. The vacuum chamber is pumped down to $5 \times 10^{-4}$Pa ~ $5 \times 10^{-1}$Pa and then back filled with a working gas such as Ar gas to reach a working pressure of 5Pa ~ 60Pa. The working voltages are applied to the three grids. Under the action of the PLUME effect and the hollow cathode, the infiltrated elements in the S pole and P pole made of solid materials are in the form of ions, atoms, and particle clusters. which are sputtered out and moved to the working surface of W-pole, attracting to the surface of the workpiece. Due to the high temperature of the surface of the workpiece, a multi-layer coating, including deposition, diffusion, and the diffusion layers can be obtained.

## 11. Double Glow Plasma Chemistry

In 1874, P. A. Thenard was the first one successfully obtained organic thin film in gas discharge. In 1905, J. N. Collie made the ethylene to be liquid synthesis in gas discharge. A large number of experimental results have proved that many chemical reactions that cannot be achieved under atmospheric conditions can be achieved under glow discharge conditions. Some chemical reactions can be carried out only at high temperatures but the reaction can be realized at low temperatures under glow discharge conditions. In 1967, F. K. Mctaggart proposed a concept of "Plasma Chemistry" to classify chemistry reactions in the gas discharge [51]. Nowadays, with the development of vacuum technology and gas discharge, plasma chemistry has made great progress.

It is expected that the phenomenon of Double Glow Discharge will be applicable in plasma chemistry. Its application can strengthen the glow discharge and enhance its controllability, and also to supply metallic elements in the glow discharge for the chemical reaction.

## 12. Application of Double Glow Plasma Technology in Diamond Film

Diamond films have the characteristics of chemical stability, super hardness, abrasion resistance, corrosion resistance, insulation and high light transmittance and refractive index. They have wide application prospects in the fields of machinery, electronics, optics, acoustics and medical devices. Diamond films can be synthetized by low pressure or atmospheric pressure chemical vapor deposition (CVD) [52]. The preparation methods of diamond films include thermochemical vapor deposition (TCVD) and plasma chemical vapor deposition (PCVD). Diamond-like carbon (DLC) film is similar to the diamond film [53]. The preparation of DLC membranes has developed rapidly with a variety of preparation methods. These methods can be classified into two categories: physical vapor deposition and chemical vapor deposition. However, the adhesion strength between the diamond coatings and the bonding substrates (e.g. metals or alloys) has limited the application of diamond film due to the chemical inertness of diamond and the poor wettability of molten metal.

By using double cathode hollow cathode discharge, the working carrier gas of hydrocarbon is activated and dissociated to form a high density plasma, which has many efforts in deposit on diamond film or surface modification for the diamond film. The Xu-Tec can be employed as the metallization method of preparing W metallic coatings and $Ta_xC$ interlayer onto CVD diamond films [54]. In recent years, the application of Xu-Tec process in diamond has following three aspects:

(1) Surface metallurgy on diamond films: the diamond film cannot directly be formed on the surface of iron and steel materials, but tungsten can be formed on the surface of iron and steel materials by the double glow technology. In the previous research [55-56], the continuous and compact W metallic coatings were formed between W-metalized diamond films and commercial cemented carbide (WC-Co) inserts. During the metalizing process, WC and $W_2C$ were formed by the inter-diffusion and reaction between W and C. The as-prepared W coatings were continuous and compact and WC and W2C phases were formed by the inter-diffusion and reaction between W and C. The cross section of W coatings was smooth and compact, which was made up of clusters and slender fibrous crystals, as shown in Figure 16. In addition, the W-diamond interface exhibits no cracks, which indicate a well-bonded interface.

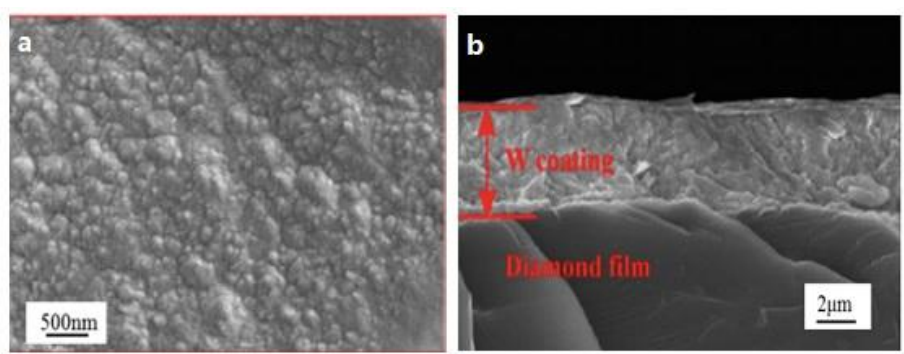

Fig 16 (a) Surface morphologies of W-alloy on CVD diamond film and (b) Cross-sectional morphology of the as-deposited W metallic coating on a CVD diamond

(2) Preparing diamond coating with the pre-deposition interlayer on cemented carbide (WC-Co) substrate: although diamond coating can significantly increase the life of cemented carbide (WC-Co) tools applied in cutting difficult-to-cut materials, the deteriorated adhesion problem, caused by the deleterious catalytic properties of cobalt (Co) binder phase during diamond growth process, limits the applications of these tools. The pre-deposition of $Ta_xC$ [56], HfC, SiC/HfC interlayers and HfC-SiC/HfC double-interlayer [57-59] on cemented carbide (WC-Co) substrate can improve the adhesion of diamond coating by double glow plasma surface alloying technique. $Ta_xC$ interlayer with an inner diffusion layer and an outer deposition layer composed of $Ta_2C$ and TaC nanocrystalline, exhibited a special compact surface morphology formed of flower-shaped pits as shown in Figure 17. As the gradual element distributions existed in the diffusion layer, the interlayer displayed a superior adherence to the substrate with significantly enhanced surface microhardness to the original substrate. HfC, SiC and its double-interlayer can also exhibit the similar the gradual element distributions, shown in Figure 18. The study indicated that both TaxC diffusion-deposition interlayer and HfC-SiC/HfC double-interlayer can be promising ways to improve the performance of diamond coatings used in cutting tools.

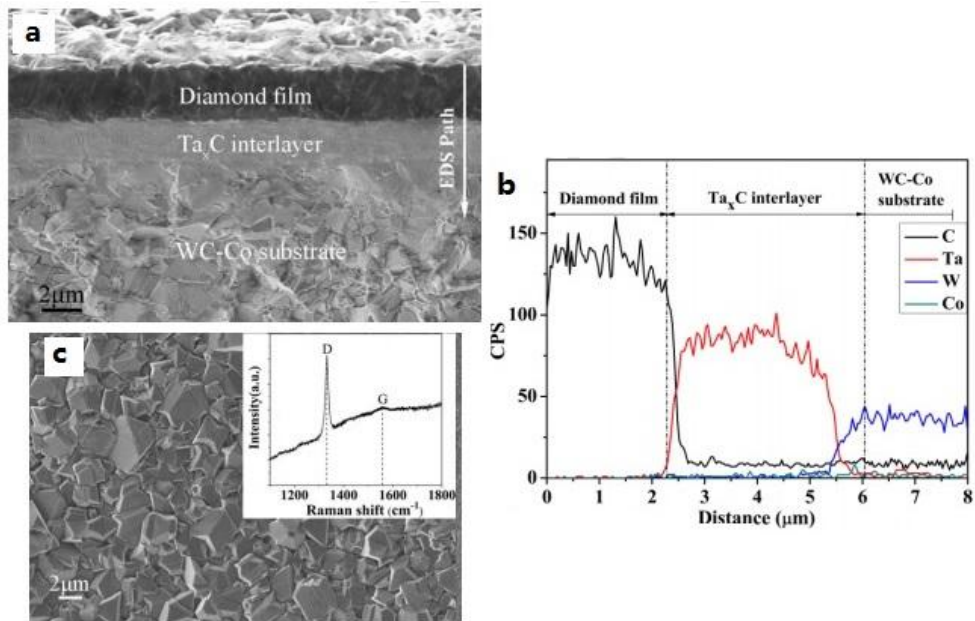

FIG. 17. (a) Cross-section morphology, (b) elemental distributions of the diamond SEM image and (c) surface morphology and Raman spectrum of the diamond coating deposited on the interlayered substrates

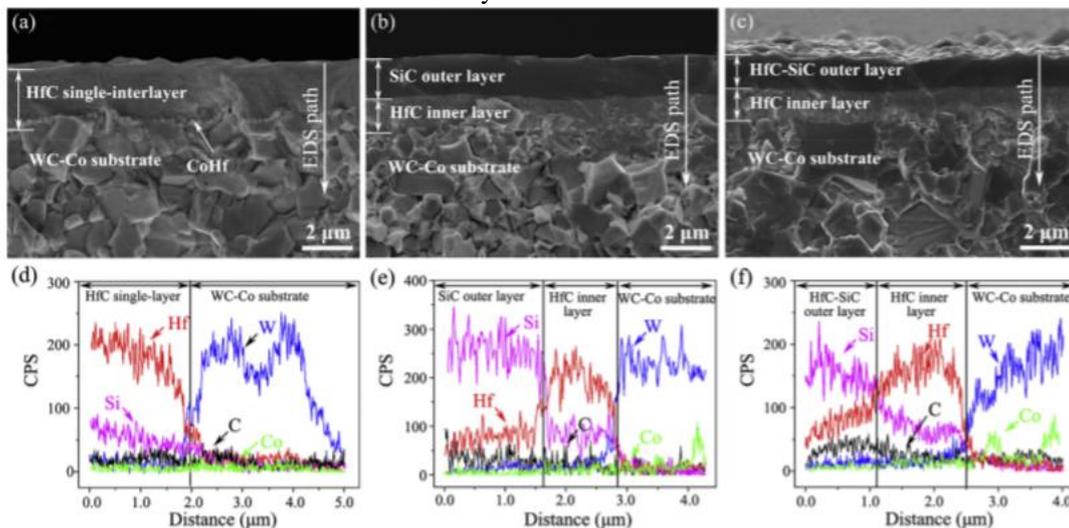

FIG. 18. The cross-section SEM images and elemental distributions of the as-prepared samples: (a, d) HfC single-interlayer, (b, e) SiC/HfC double-interlayer, and (c, f) HfC-SiC/HfC double-interlayer.

The modification method for the diamond coating by double glow hollow cathode discharge has the following advantages:

(1) Improve the grip force of diamond. The as-prepared W-metalized diamond film treated at 800 ℃ possesses the maximums in the load and the strength are approximately 17.1 N and 249 MPa, respectively.

(2) Realize the welding between diamond and diamond or diamond and metal

material.

(3) The double glow plasma surface alloying technique is an efficient method for preparing the interface between common metals and diamond coatings by forming gradient alloying layer, which can relax the thermal stresses by a diffusion layer with micron thickness.

## 14. Summery and Prospect

Plasma, as the fourth form of matter, has been widely used in the development of new technologies and the preparation of new materials. Surface sputtering cleaning, surface activation, surface etching and surface coating of solid materials by plasma have been widely applied in semiconductor chip processing and surface modification of industrial solid materials. Its social benefits and economic value are immeasurable. However, the application of plasma in metal materials and machinery manufacturing industry is still at its infancy except plasma nitriding before the advent of the Xu-Tec technology.

(1) Based on double glow discharge and inspired by double glow plasma surface metallurgy technology, we invented a series of technologies, such as the double glow plasma grapheme technology, double glow plasma brazing technology, double glow plasma sintering technology, double glow plasma nanotechnology, double glow plus high frequency plasma surface alloying, tri-glow plasma surface alloying and so on.

(2) The Double Glow Discharge Phenomenon is the origin of a series innovative technologies. It is a new discovery with important scientific significance. We believe it will have more application potential in other technology areas in the future.

(3) Except the Xu-Tec technology, most of the technologies introduced in this paper have not been studied in depth yet. We need further research and application development.

(4) We hope plasma technologies introduced in this paper will promote the development and applications of plasma in metal materials and mechanical manufacturing; and even will gradually expand to form a new field of science and technology characterized by plasma.

## Acknowledgment

The authors wish to thank their colleagues and all doctorate students for their contributions in the Xu-Tec process.
Thank Dr. Siu-Ping Hong for reviewing and carefully modifying the text.